\pdfoutput=1
\documentclass[12pt,a4paper]{article}

\usepackage{ifthen} 
\newboolean{pdflatex}
\setboolean{pdflatex}{true} 

\newboolean{articletitles}
\setboolean{articletitles}{true} 

\newboolean{uprightparticles}
\setboolean{uprightparticles}{false} 

\def\paperauthors{LHCb collaboration} 
\def\paperasciititle{Search for time-dependent CP violation in D0 -> pi+ pi- pi0 decays} 
\def\papertitle{Search for time-dependent {{\ensuremath{C\!P}}\xspace} violation in $D^0 \rightarrow \pi^+ \pi^- \pi^0$ decays} 
\def\paperkeywords{{High Energy Physics}, {LHCb}} 
\def\papercopyright{\the\year\ CERN for the benefit of the LHCb collaboration} 
\def\paperlicence{CC BY 4.0 licence}
\def\paperlicenceurl{https://creativecommons.org/licenses/by/4.0/}

\usepackage[top=1in, bottom=1.25in, left=1in, right=1in]{geometry}

\usepackage{microtype}
\usepackage{lineno}  
\usepackage{xspace} 
\usepackage{caption} 

\usepackage{graphicx}  
\usepackage{color}
\usepackage{colortbl}
\graphicspath{{./figs/}} 

\usepackage{amsmath} 
\usepackage{amssymb}
\usepackage{amsfonts}
\usepackage{upgreek} 

\newcommand*\patchAmsMathEnvironmentForLineno[1]{%
\expandafter\let\csname old#1\expandafter\endcsname\csname #1\endcsname
\expandafter\let\csname oldend#1\expandafter\endcsname\csname
end#1\endcsname
 \renewenvironment{#1}%
   {\linenomath\csname old#1\endcsname}%
   {\csname oldend#1\endcsname\endlinenomath}%
}
\newcommand*\patchBothAmsMathEnvironmentsForLineno[1]{%
  \patchAmsMathEnvironmentForLineno{#1}%
  \patchAmsMathEnvironmentForLineno{#1*}%
}
\AtBeginDocument{%
\patchBothAmsMathEnvironmentsForLineno{equation}%
\patchBothAmsMathEnvironmentsForLineno{align}%
\patchBothAmsMathEnvironmentsForLineno{flalign}%
\patchBothAmsMathEnvironmentsForLineno{alignat}%
\patchBothAmsMathEnvironmentsForLineno{gather}%
\patchBothAmsMathEnvironmentsForLineno{multline}%
\patchBothAmsMathEnvironmentsForLineno{eqnarray}%
}

\usepackage{hyperxmp}

\usepackage[pdftex,
            pdfauthor={\paperauthors},
            pdftitle={\paperasciititle},
            pdfkeywords={\paperkeywords},
            pdfcopyright={Copyright (C) \papercopyright},
            pdflicenseurl={\paperlicenceurl}]{hyperref}

\usepackage[colorinlistoftodos,textsize=scriptsize]{todonotes}

\usepackage[bottom,flushmargin,hang,multiple]{footmisc}

\usepackage[all]{hypcap} 

\usepackage{xspace} 
\usepackage{upgreek}

\def\lhcb   {\mbox{LHCb}\xspace}

\def\babar  {\mbox{BaBar}\xspace}
\def\belle  {\mbox{Belle}\xspace}

\def\cleo   {\mbox{CLEO}\xspace}
\def\cdf    {\mbox{CDF}\xspace}

\def\MagUp {\mbox{\em Mag\kern -0.05em Up}\xspace}
\def\MagDown {\mbox{\em MagDown}\xspace}

\ifthenelse{\boolean{uprightparticles}}%
{

 \def\Ppi         {\ensuremath{\uppi}\xspace}

 \def\PDelta      {\ensuremath{\Delta}\xspace}                 
 \def\PXi         {\ensuremath{\Xi}\xspace}                 
 \def\PLambda     {\ensuremath{\Lambda}\xspace}                 
 \def\PSigma      {\ensuremath{\Sigma}\xspace}                 
 \def\POmega      {\ensuremath{\Omega}\xspace}                 
 \def\PUpsilon    {\ensuremath{\Upsilon}\xspace}
 \let\oldPi\Pi
 \def\PPi         {\ensuremath{\oldPi}\xspace}

 \def\PB      {\ensuremath{\mathrm{B}}\xspace}                 
                  
 \def\PD      {\ensuremath{\mathrm{D}}\xspace}

 \def\PK      {\ensuremath{\mathrm{K}}\xspace}

 \def\Pb      {\ensuremath{\mathrm{b}}\xspace}                 
 \def\Pc      {\ensuremath{\mathrm{c}}\xspace}

 \def\Pi      {\ensuremath{\mathrm{i}}\xspace}

 \def\Pp      {\ensuremath{\mathrm{p}}\xspace}

 \def\Ps      {\ensuremath{\mathrm{s}}\xspace}

 \def\thebaroffset{0.0em}
}
{

 \def\Ppi         {\ensuremath{\pi}\xspace}

 \mathchardef\PDelta="7101
 \mathchardef\PXi="7104
 \mathchardef\PLambda="7103
 \mathchardef\PSigma="7106
 \mathchardef\POmega="710A
 \mathchardef\PUpsilon="7107
 \mathchardef\PPi="7105
                  
 \def\PB      {\ensuremath{B}\xspace}                 
                  
 \def\PD      {\ensuremath{D}\xspace}

 \def\PK      {\ensuremath{K}\xspace}

 \def\Pb      {\ensuremath{b}\xspace}                 
 \def\Pc      {\ensuremath{c}\xspace}

 \def\Pi      {\ensuremath{i}\xspace}

 \def\Pp      {\ensuremath{p}\xspace}

 \def\Ps      {\ensuremath{s}\xspace}

 \def\thebaroffset{0.18em}
}
\newcommand{\offsetoverline}[2][\thebaroffset]{\kern #1\overline{\kern -#1 #2}}%

\makeatletter
\ifcase \@ptsize \relax
  \newcommand{\miniscule}{\@setfontsize\miniscule{4}{5}}
\or
  \newcommand{\miniscule}{\@setfontsize\miniscule{5}{6}}
\or
  \newcommand{\miniscule}{\@setfontsize\miniscule{5}{6}}
\fi
\makeatother

\DeclareRobustCommand{\optbar}[1]{\shortstack{{\miniscule (\rule[.5ex]{1.25em}{.18mm})}
  \\ [-.7ex] $#1$}}

\def\squark    {{\ensuremath{\Ps}}\xspace}

\def\cquark    {{\ensuremath{\Pc}}\xspace}

\def\bquark    {{\ensuremath{\Pb}}\xspace}

\def\pion   {{\ensuremath{\Ppi}}\xspace}
\def\piz    {{\ensuremath{\pion^0}}\xspace}
\def\pip    {{\ensuremath{\pion^+}}\xspace}
\def\pim    {{\ensuremath{\pion^-}}\xspace}

\def\kaon    {{\ensuremath{\PK}}\xspace}

\def\KorKbar {\kern \thebaroffset\optbar{\kern -\thebaroffset \PK}{}\xspace}

\def\Kp      {{\ensuremath{\kaon^+}}\xspace}
\def\Km      {{\ensuremath{\kaon^-}}\xspace}

\def\KS      {{\ensuremath{\kaon^0_{\mathrm{S}}}}\xspace}

\def\Dbar    {{\ensuremath{\offsetoverline{\PD}}}\xspace}
\def\D       {{\ensuremath{\PD}}\xspace}

\def\DorDbar {\kern \thebaroffset\optbar{\kern -\thebaroffset \PD}\xspace}
\def\Dz      {{\ensuremath{\D^0}}\xspace}
\def\Dzb     {{\ensuremath{\Dbar{}^0}}\xspace}
\def\Dp      {{\ensuremath{\D^+}}\xspace}
\def\Dm      {{\ensuremath{\D^-}}\xspace}

\def\DpDm    {\ensuremath{\Dp {\kern -0.16em \Dm}}\xspace}

\def\Dstarp  {{\ensuremath{\D^{*+}}}\xspace}

\def\Dstarpm {{\ensuremath{\D^{*\pm}}}\xspace}

\def\theDstarp{{\ensuremath{\D^{*}(2010)^{+}}}\xspace}

\def\B       {{\ensuremath{\PB}}\xspace}

\def\BorBbar {\kern \thebaroffset\optbar{\kern -\thebaroffset \PB}\xspace}

\def\Bd      {{\ensuremath{\B^0}}\xspace}

\def\BdorBdbar {\kern \thebaroffset\optbar{\kern -\thebaroffset \Bd}\xspace}

\def\Bs      {{\ensuremath{\B^0_\squark}}\xspace}

\def\BsorBsbar {\kern \thebaroffset\optbar{\kern -\thebaroffset \Bs}\xspace}

\def\Y#1S{\ensuremath{\PUpsilon{(#1S)}}\xspace}

\def\proton      {{\ensuremath{\Pp}}\xspace}

\def\LorLbar     {\kern \thebaroffset\optbar{\kern -\thebaroffset \PLambda}\xspace}

\newcommand{\decay}[2]{\ensuremath{#1\!\to #2}\xspace} 

\def\to                 {\ensuremath{\rightarrow}\xspace}

\def\CP                {{\ensuremath{C\!P}}\xspace}

\newcommand{\dm}{{\ensuremath{\Delta m}}\xspace}

\def\AT#1     {\ensuremath{A_{\mathrm{T}}^{#1}}\xspace}           

\def\C#1      {\ensuremath{\mathcal{C}_{#1}}\xspace}                       
\def\Cp#1     {\ensuremath{\mathcal{C}_{#1}^{'}}\xspace}                    
\def\Ceff#1   {\ensuremath{\mathcal{C}_{#1}^{\mathrm{(eff)}}}\xspace}        
\def\Cpeff#1  {\ensuremath{\mathcal{C}_{#1}^{'\mathrm{(eff)}}}\xspace}       
\def\Ope#1    {\ensuremath{\mathcal{O}_{#1}}\xspace}                       
\def\Opep#1   {\ensuremath{\mathcal{O}_{#1}^{'}}\xspace}                    

\newcommand{\ket}[1]{\ensuremath{|#1\rangle}}              

\newcommand{\aunit}[1]{\ensuremath{\text{\,#1}}}       

\newcommand{\tev}{\aunit{Te\kern -0.1em V}\xspace}
\newcommand{\gev}{\aunit{Ge\kern -0.1em V}\xspace}
\newcommand{\mev}{\aunit{Me\kern -0.1em V}\xspace}
\newcommand{\kev}{\aunit{ke\kern -0.1em V}\xspace}
\newcommand{\ev}{\aunit{e\kern -0.1em V}\xspace}
 
\newcommand{\mevc}{\ensuremath{\aunit{Me\kern -0.1em V\!/}c}\xspace}
\newcommand{\gevc}{\ensuremath{\aunit{Ge\kern -0.1em V\!/}c}\xspace}
\newcommand{\mevcc}{\ensuremath{\aunit{Me\kern -0.1em V\!/}c^2}\xspace}
\newcommand{\gevcc}{\ensuremath{\aunit{Ge\kern -0.1em V\!/}c^2}\xspace}

\def\fb   {\ensuremath{\aunit{fb}}\xspace}
\def\invfb   {\ensuremath{\fb^{-1}}\xspace}

\def\fs   {\aunit{fs}}

\newcommand{\chisq}{\ensuremath{\chi^2}\xspace}

\newcommand{\chisqip}{\ensuremath{\chi^2_{\text{IP}}}\xspace}

\def\gsim{{~\raise.15em\hbox{$>$}\kern-.85em
          \lower.35em\hbox{$\sim$}~}\xspace}
\def\lsim{{~\raise.15em\hbox{$<$}\kern-.85em
          \lower.35em\hbox{$\sim$}~}\xspace}

\def\sqs   {\ensuremath{\protect\sqrt{s}}\xspace}

\def\roofit     {\mbox{\textsc{RooFit}}\xspace}

\def\tell1  {TELL1\xspace}
\def\ukl1   {UKL1\xspace}

\newcommand{\lhcborcid}[1]{\href{https://orcid.org/#1}{\hspace*{0.1em}\raisebox{-0.45ex}{\includegraphics[width=1em]{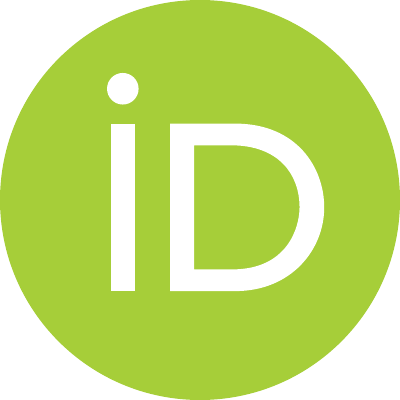}}}}

\usepackage{cite} 
\usepackage{mciteplus}

\usepackage{longtable} 

\usepackage{subcaption}

\begin{document}

\renewcommand{\thefootnote}{\fnsymbol{footnote}}
\setcounter{footnote}{1}


\begin{titlepage}
\pagenumbering{roman}

\vspace*{-1.5cm}
\centerline{\large EUROPEAN ORGANIZATION FOR NUCLEAR RESEARCH (CERN)}
\vspace*{1.5cm}
\noindent
\begin{tabular*}{\linewidth}{lc@{\extracolsep{\fill}}r@{\extracolsep{0pt}}}
\ifthenelse{\boolean{pdflatex}}
{\vspace*{-1.5cm}\mbox{\!\!\!\includegraphics[width=.14\textwidth]{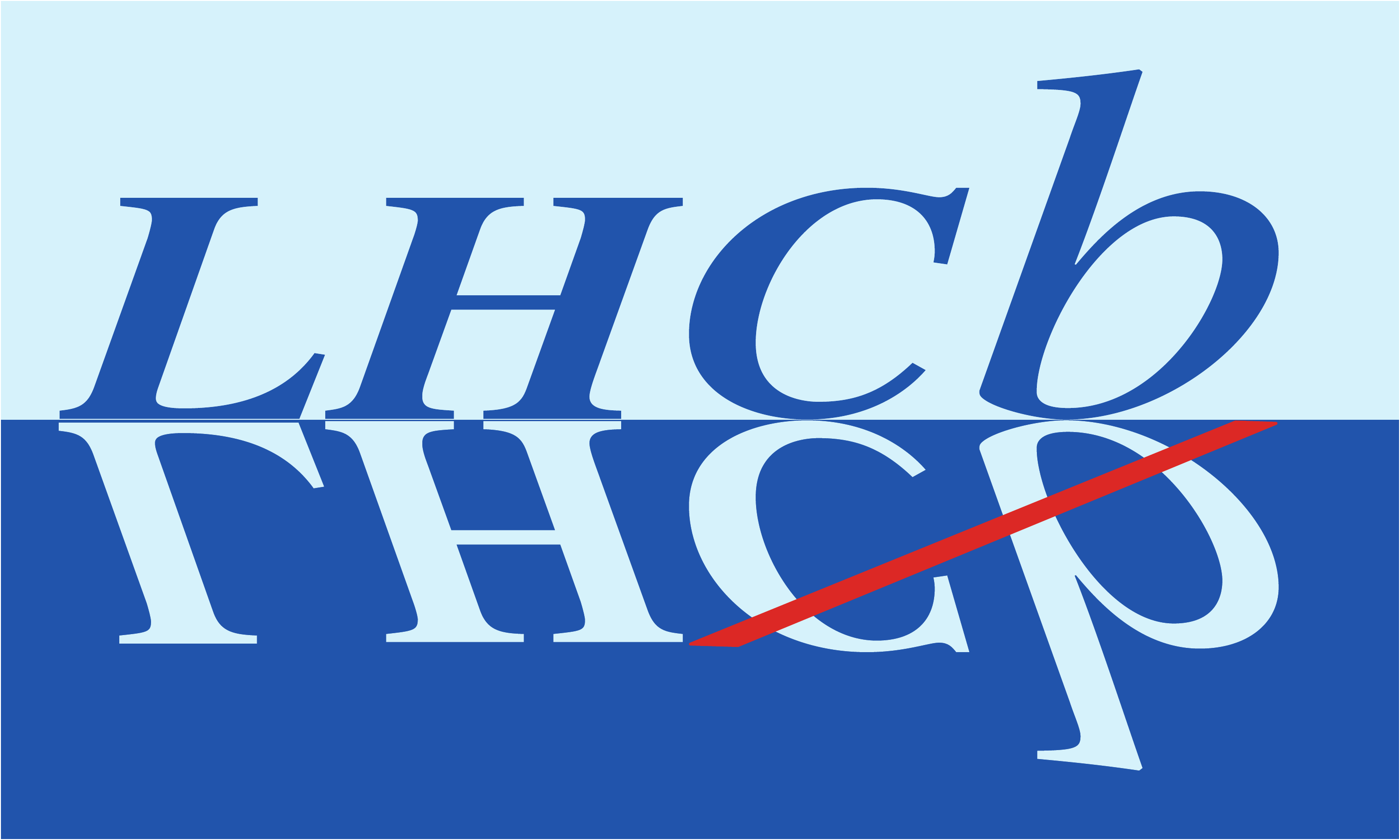}} & &}%
{\vspace*{-1.2cm}\mbox{\!\!\!\includegraphics[width=.12\textwidth]{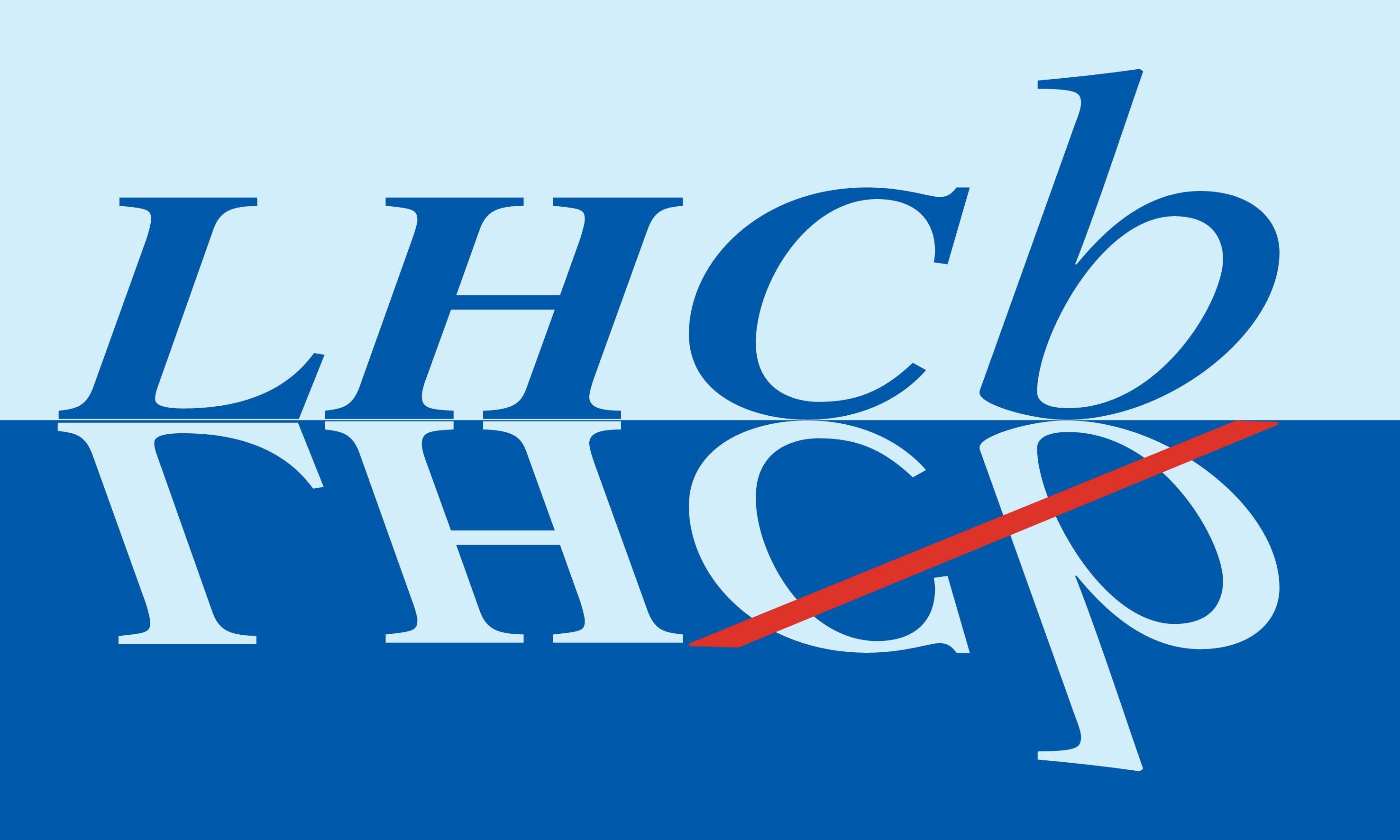}} & &}%
\\
 & & CERN-EP-2024-111 \\  
 & & LHCb-PAPER-2024-003 \\  
 & & September 9, 2024\\
 & & \\
\end{tabular*}

\vspace*{4.0cm}

{\normalfont\bfseries\boldmath\huge
\begin{center}
  \papertitle 
\end{center}
}

\vspace*{2.0cm}

\begin{center}
\paperauthors\footnote{Authors are listed at the end of this Letter.}
\end{center}

\vspace{\fill}

\begin{abstract}
  \noindent A measurement of time-dependent {{\ensuremath{C\!P}}\xspace} violation in $D^0 \rightarrow \pi^+ \pi^- \pi^0$ decays using a $pp$ collision data sample collected by the {\mbox{LHCb}\xspace} experiment in 2012 and from 2015 to 2018, corresponding to an integrated luminosity of 7.7{\ensuremath{\fb^{-1}}\xspace}, is presented. The initial flavour of each $D^0$ candidate is determined from the charge of the pion produced in the \mbox{$D^*(2010)^+ \rightarrow D^0 \pi^+$} decay. The decay $D^0 \rightarrow K^- \pi^+ \pi^0$ is used as a control channel to validate the measurement procedure. The gradient of the time-dependent {{\ensuremath{C\!P}}\xspace} asymmetry, $\Delta Y$, in $D^0 \rightarrow \pi^+ \pi^- \pi^0$ decays is measured to be
  \begin{equation*}
      \Delta Y = (-1.3 \pm 6.3 \pm 2.4) \times 10^{-4},
  \end{equation*}
  where the first uncertainty is statistical and the second is systematic, which is compatible with {{\ensuremath{C\!P}}\xspace} conservation.
\end{abstract}

\vspace*{2.0cm}

\begin{center}
    Published in Phys. Rev. Lett. 133 (2024) 101803
\end{center}

\vspace{\fill}

{\footnotesize 
\centerline{\copyright~\papercopyright. \href{\paperlicenceurl}{\paperlicence}.}}
\vspace*{2mm}

\end{titlepage}


\newpage
\setcounter{page}{2}
\mbox{~}
%
%
%
%


\renewcommand{\thefootnote}{\arabic{footnote}}
\setcounter{footnote}{0}

\cleardoublepage


\pagestyle{plain} 
\setcounter{page}{1}
\pagenumbering{arabic}


Charge-parity (\CP) symmetry violation in the Standard Model (SM) of particle physics is insufficient to explain the observed baryon asymmetry in the visible universe~\cite{Sakharov:1967dj,Dine:2003ax, Canetti:2012zc}. This suggests the existence of \CP-violating mechanisms beyond the SM, and thus studies of \CP violation are a promising sector to probe for new physics. While \CP violation is experimentally well-established in the mesonic systems containing \bquark or \squark quarks~\cite{Christenson:1964fg, KTeV:1999kad, BaBar:2001pki, Belle:2001zzw, LHCB-PAPER-2013-018, LHCb-PAPER-2020-029}, direct \CP violation in the decay of charmed hadrons has only recently been observed in the difference between \decay{\Dz}{\Kp\Km} and \decay{\Dz}{\pip\pim} decays~\cite{LHCb-PAPER-2019-006}, and is yet to be conclusively observed in a single decay mode~\cite{LHCb-PAPER-2022-024}. This Letter presents the first measurement of time-dependent \CP violation in the singly Cabibbo-suppressed decay \decay{\Dz}{\pip\pim\piz}.\footnote{Charge-conjugate decays are implied here and throughout this Letter.}

In the \Dz meson system, the flavour eigenstates differ from the mass eigenstates. A neutral meson initially in a state of definite flavour can thus evolve in time into its antiparticle, and vice versa. The mass eigenstates are typically written in terms of the flavour eigenstates as $\ket{D_{1(2)}} = p\ket{\Dz} \mp q\ket{\Dzb}$, where $p$ and $q$ are complex numbers. The phase convention $\CP \ket{\Dz} = -\ket{\Dzb}$ is adopted with $\CP\ket{D_1} = \ket{D_1}$ in the limit of \CP conservation~\cite{HFLAV21}. The dimensionless parameters \mbox{$x \equiv (m_1 - m_2)/\Gamma$} and \mbox{$y \equiv (\Gamma_1 - \Gamma_2)/2\Gamma$},\footnote{Natural units with $\hbar = c = 1$ are used throughout.} where $m_{1(2)}$ and $\Gamma_{1(2)}$ are the mass and width of the $\ket{D_{1(2)}}$ mass eigenstate and \mbox{$\Gamma = (\Gamma_1 + \Gamma_2)/2$} is the mean width, define the mixing properties. Defining $A_f$ ($\bar{A}_f$) as the amplitude of the decay \decay{\Dz}{f} (\decay{\Dzb}{f}), direct \CP violation occurs if $|A_f/\bar{A}_f| \neq 1$ for a self-conjugate final state $f$. Time-dependent \CP violation occurs in mixing if $|q/p| \neq 1$, and in the interference of mixing and decay when $\phi_f \neq 0$, where $\phi_f = \arg(-q\bar{A}_f/pA_f)$. Since the mixing parameters $x$ and $y$ are both $< 1\%$ in the charm system and \CP-violation effects are small~\cite{HFLAV21, LHCb-PAPER-2016-032, GammaCombo}, the time-dependent asymmetry for a \Dz meson decaying to a \CP eigenstate, $f_\CP$, can be expanded to first order in the $D^0$ decay time, $t$, as
\begin{equation}
    \begin{aligned}
        A_{\CP}(f_\CP, t) &\equiv \frac{\Gamma_{\decay{\Dz}{f_\CP}}(t) - \Gamma_{\decay{\Dzb}{f_\CP}}(t)}{\Gamma_{\decay{\Dz}{f_\CP}}(t) + \Gamma_{\decay{\Dzb}{f_\CP}}(t)} \\
        &\approx a_{f_\CP}^\text{dir} + \Delta Y_{f_\CP} \frac{t}{\tau_\Dz}.
    \end{aligned}
\label{eq:acp_t}
\end{equation}
The constant term $a_{f_\CP}^\text{dir}$ arises from direct \CP violation, $\tau_\Dz = 410.3 \pm 1.0 \fs$~\cite{PDG2022, Belle-II:2021cxx} is the \Dz meson lifetime, and $\Gamma_{\decay{\Dz}{f_\CP}}(t)$ ($\Gamma_{\decay{\Dzb}{f_\CP}}(t)$) is the time-dependent decay rate of a \mbox{\Dz (\Dzb)} meson to the final state $f_{\CP}$. Neglecting direct \CP violation, the size of the gradient $\Delta Y_{f_\CP}$ becomes independent of the final state and can be approximately expressed in terms of the underlying mixing and \CP violation parameters as~\cite{Kagan:2020vri}
\begin{equation} 
    \Delta Y_{f_\CP} \approx 
    \frac{\eta_{f_{\CP}}}{2}
    \left[
        \left(\left|\frac{q}{p}\right| + \left|\frac{p}{q}\right| \right)\,x\sin\phi
        - \left(\left|\frac{q}{p}\right| - \left|\frac{p}{q}\right| \right)\,y\cos\phi
    \right],
\end{equation}
where $\eta_{f_\CP}$ is the \CP eigenvalue of the final state and $\phi = \arg(q/p)$. Thus, the gradient can be defined in terms of the universal \CP-violating parameter $\Delta Y \equiv \eta_{f_{\CP}} \Delta Y_{f_{\CP}}$.

The parameter $\Delta Y$ has been measured in the two-body decay modes \decay{\Dz}{\pip\pim} and \decay{\Dz}{\Km\Kp} by the \babar~\cite{BaBar:2012bho}, \cdf~\cite{CDF:2014wyb}, \belle~\cite{Belle:2015etc} and \lhcb~\cite{LHCb-PAPER-2014-069, LHCb-PAPER-2016-063, LHCb-PAPER-2019-032, LHCb-PAPER-2020-045} collaborations. Assuming universality of $\Delta Y$ across all decay modes, the current world average is \mbox{$\Delta Y = (0.9 \pm 1.1) \times 10^{-4}$}~\cite{HFLAV21}. The \lhcb collaboration has also measured the parameter $\Delta y = - \Delta Y$~\cite{Kagan:2020vri} in \decay{\Dz}{\KS\pip\pim} decays~\cite{LHCb-PAPER-2021-009, LHCb-PAPER-2022-020}.

Phase-space integrated analyses of multi-body \Dz decays have diluted sensitivity to $\Delta Y$ due to a mixture of \CP-even and \CP-odd contributions from intermediate states. The effective gradient of the time-dependent asymmetry is given by~\cite{Malde:2015xra}
\begin{equation}
    \Delta Y_{f}^\text{eff} = (2F_+^{f} - 1)\Delta Y,
\end{equation}
where $F_+^f$ is the \CP-even fraction of the decay \decay{\Dz}{f}. For the decay \decay{\Dz}{\pip\pim\piz}, the \CP-even fraction has been measured using \cleo-$c$ data to be \mbox{$F^{\pi\pi\pi}_+ = 0.973 \pm 0.017$}~\cite{Malde:2015mha}, providing almost undiluted sensitivity.

The Cabibbo-favoured decay \decay{\Dz}{\Km\pip\piz} is used to validate the analysis procedure. Since the decay \decay{\Dzb}{\Km\pip\piz} is doubly Cabibbo-suppressed, the final state \Km\pip\piz is almost flavour specific. Therefore, \CP-violation effects in mixing and in the interference of mixing and decay are suppressed with respect to the signal channel. The corresponding \CP-violating parameter in this mode, $\Delta Y_{K\pi\pi}$, has been estimated from the most recent world averages of the relevant parameters~\cite{HFLAV21, PDG2022, LHCb-PAPER-2016-032} to be $|\Delta Y_{K\pi\pi}| < 2.5 \times 10^{-5}$ at 90\% confidence level.

Samples of \decay{\Dz}{\pip\pim\piz} decays are reconstructed from proton-proton (\proton\proton) collisions collected by the \lhcb experiment in 2012 and from 2015 to 2018, corresponding to an integrated luminosity of 7.7\invfb at centre-of-mass energies of $\sqs = 8$ and $13\tev$, respectively. Throughout this Letter, the data sample collected in 2012 (2015--18) will be referred to as the Run 1 (2) sample. The flavour of the \Dz meson at production is inferred from the charge of the pion produced in the preceding \decay{\theDstarp}{\Dz\pip} decay. The \theDstarp meson and the pion originating from its decay will be referred to as the \Dstarp meson and tag pion throughout this Letter, respectively. 

The LHCb detector~\cite{LHCb-DP-2008-001,LHCb-DP-2014-002} is a single-arm forward spectrometer covering the pseudorapidity range $2 < \eta < 5$, designed for the study of particles containing \bquark\ or \cquark\ quarks. The detector elements that are particularly relevant to this analysis are: a silicon-strip vertex detector surrounding the $pp$ interaction region that allows a precise measurement of the flight distance of $D^0$ mesons; a tracking system that provides a measurement of the momentum of the charged particles; a set of ring-imaging Cherenkov detectors that provide charged particle identification; and a calorimeter system which identifies photons and provides a measurement of their energy. The flight distance of the \Dz meson is defined as the distance between its measured decay vertex and the primary \proton\proton interaction vertex (PV). Neutral pions are reconstructed in their diphoton decay. If the energy clusters produced by the two photons in the calorimeter overlap, the candidate is labelled as {\it merged}; otherwise, the candidate is labelled as {\it resolved}.

Online event selection is performed by a trigger consisting of a hardware stage, using information from the calorimeter and muon systems, followed by two software stages. In this analysis, no explicit requirements are imposed at the hardware level, since all hardware selections are found to contribute to the sample~\cite{LHCb-PAPER-2014-054}. At the first software stage in the Run 1 running period, at least one of the charged pions from the \Dz meson decay must satisfy criteria on its momentum, \chisqip and track quality. The quantity \chisqip\ is defined as the difference in the vertex-fit \chisq of a PV reconstructed with and without the given track or particle under consideration. For the Run 2 sample either one of the pions from the \Dz decay must satisfy a similar set of criteria, or both of the tracks must satisfy a set of looser criteria and their combined vertex must satisfy additional vertex-quality requirements. At the second software stage, a full or partial reconstruction of the \Dstarp decay is performed. In the full reconstruction, pairs of oppositely-charged pion candidates that form a good-quality vertex are combined. A neutral pion candidate is added to form a \Dz candidate. Finally, the \Dz candidate is combined with a low-momentum charged pion candidate to reconstruct a \Dstarp candidate. In the partial reconstruction, the \piz candidate is not considered in the trigger. Additional selection criteria are imposed based on momenta, track and vertex quality, and displacement from the PV. For the Run 1 sample, only partially reconstructed candidates are considered with a cut-based selection. For the Run 2 sample, fully reconstructed candidates must satisfy a cut-based selection, while partially reconstructed candidates must fulfil a requirement on the output of a bonsai-Boosted Decision Tree (BDT) classifier~\cite{BBDT}.

\begin{figure}[tb]
  \begin{center}
    \includegraphics[width=0.49\linewidth]{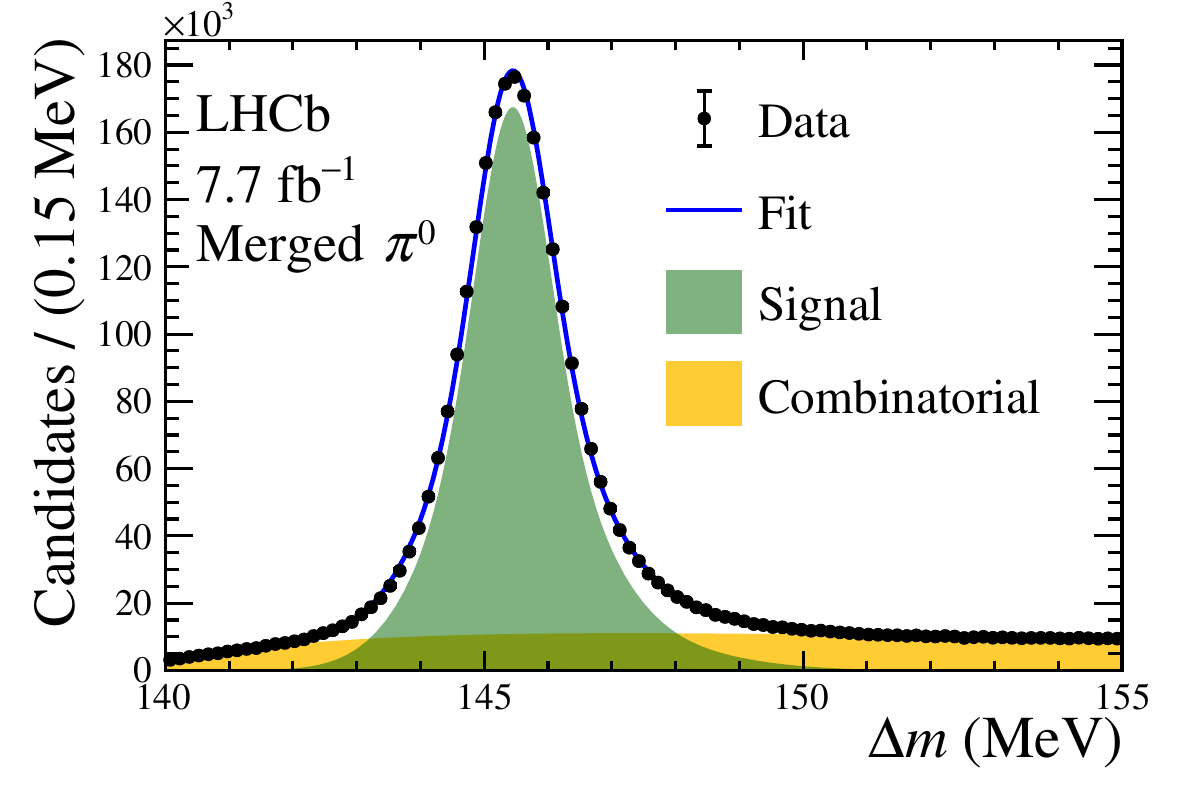}
    \includegraphics[width=0.49\linewidth]{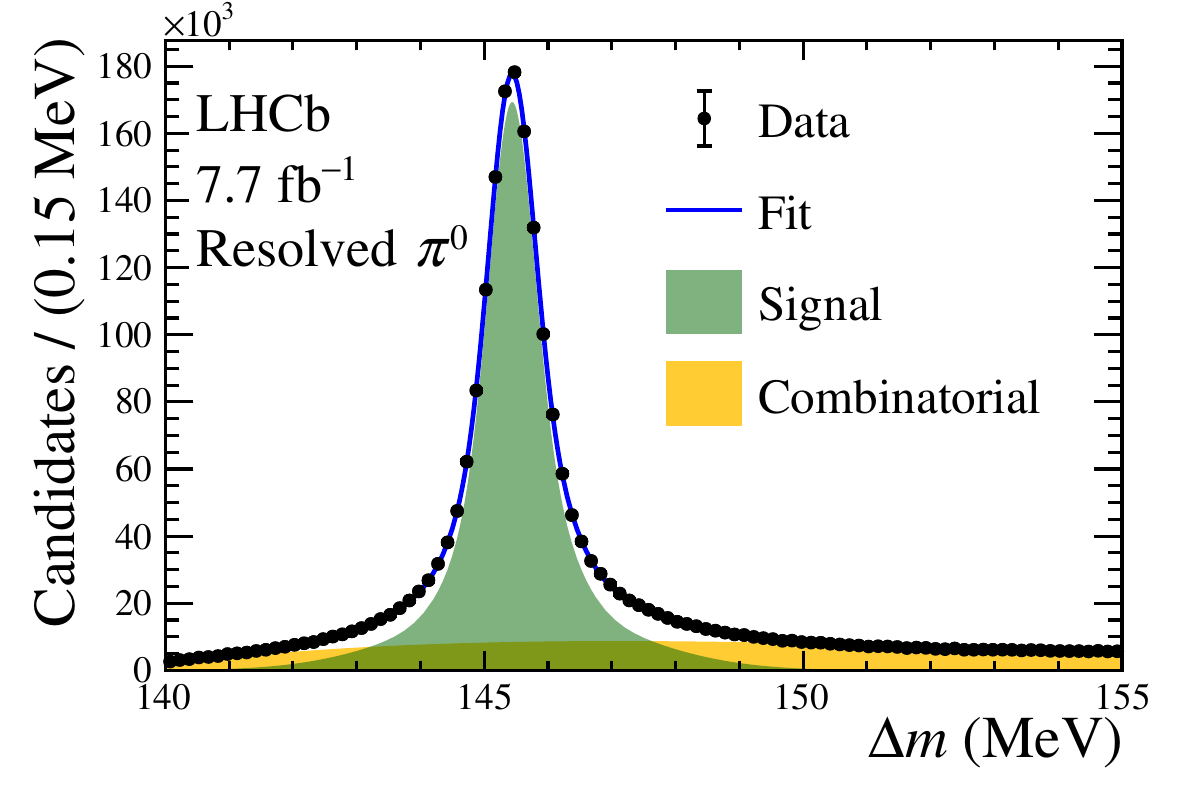}
    \vspace*{-0.5cm}
  \end{center}
  \caption{\small Distributions of $\dm$ in the signal channel after all selection requirements for the (left) merged and (right) resolved $\pi^0$ categories. The fit used to evaluate the signal yields is also shown.}
  \label{fig:deltam}
\end{figure}

Offline, additional requirements are imposed on kinematic and particle-identification observables. Secondary decays --- where the \Dstarp meson is produced in the decay of a \bquark hadron rather than in the initial \proton\proton collision --- are suppressed by imposing tight requirements on the measured flight distance of the \Dstarp candidate and the \chisqip of the \Dz candidate. The mass of the pair of charged pions from the \Dz decay is required to be inconsistent with the known \KS mass~\cite{PDG2022} to suppress the background from \decay{\Dz}{\KS\piz} decays. The background from \decay{\Dz}{\Km\pip\piz} decays, where the \Km meson is misidentified as a pion, is suppressed by requiring \Dz candidates to have a reconstructed mass, $m(\Dz)$, close to the known value, $m_\text{PDG}(\Dz)$~\cite{PDG2022}. Candidates in the resolved category must satisfy $|m(\Dz) - m_\text{PDG}(\Dz)| < 60\mev$ and candidates in the merged sample must satisfy $-60 < m(\Dz) - m_\text{PDG}(\Dz) < 120\mev$. Other potential sources of misidentified and misreconstructed backgrounds are found to be negligible. The same requirements are applied to candidates in the control channel, with the exceptions of the particle-identification requirements and the \KS meson veto. The kinematics of each decay chain are fitted~\cite{Hulsbergen:2005pu} with the constraints that the \Dstarp candidate must originate from a PV and the \piz candidate has its known mass~\cite{PDG2022}. The combinatorial background is further suppressed by a set of BDT classifiers trained on the data samples, separately for the merged and resolved \piz categories and for the Run 1 and Run 2 samples. Over-training effects are controlled by training the classifiers on only around \mbox{25\% (6\%)} of the data in the \mbox{Run 1 (2)} sample for each \piz reconstruction category, and monitoring the classifiers' performances on equivalently sized testing samples. The same BDT classifiers are applied to the signal and control channels, with different requirements on the output. The selections on the BDT classifier outputs are optimised by maximising the metric $N/\sigma_N$, where $N$ is the number of signal decays and $\sigma_N$ is its uncertainty obtained from a fit to the distribution of the \Dstarp and \Dz mass difference,  \mbox{$\dm \equiv m(\Dstarp) - m(\Dz)$}. Finally, when more than one candidate originates from a single event in the final sample, only one is randomly selected. Figure~\ref{fig:deltam} shows the distributions of  \dm for all selected candidates in the signal channel, separately for the two \piz reconstruction categories. A sum of three Gaussian distributions is used to model the signal shape. The combinatorial background is modelled by the empirical \textsc{RooDstD0BG} function from the \roofit software package~\cite{ROOT, RooFit}. Signal yields of 2.3 (18) and 1.5 (20) million are obtained in the signal (control) channel for the merged and resolved categories, respectively.

The data are analysed in subsamples split by data-taking year, magnet polarity and \piz reconstruction category. In order to avoid experimenter’s bias, the results of the analysis were not examined until the full procedure was finalised.

Each data sample is divided into 21 --- approximately evenly populated --- $\Dz$ decay-time bins in the range $[0.6, 8.0]\, \tau_\Dz$. The decay time of each \Dz candidate is calculated as the product of the known \Dz mass and its measured flight distance, divided by its measured momentum. The asymmetry in each bin is determined from a binned maximum-likelihood fit to the \dm distributions of selected \Dz and \Dzb candidates. Determining the yields by fitting the \dm distribution --- rather than the $m(\Dz)$ distribution --- removes background from real \Dz candidates combined with an erroneously tagged pion. The fits are performed simultaneously to \Dz and \Dzb candidates, with shared shape parameters but separate yields for the signal and background components. The measured asymmetry is defined as
\begin{equation}
    A_\text{meas}(\langle t/\tau_\Dz \rangle_i) \equiv \frac{N_\Dz^i - N_\Dzb^i}{N_\Dz^i + N_\Dzb^i},
\end{equation}
where $N_\Dz^i$ and $N_\Dzb^i$ are the signal yields of \Dz and \Dzb decays in a given decay-time bin, $i$, respectively. The mean decay time in each bin, $\langle t/\tau_\Dz \rangle_i$, is calculated as a weighted average of the decay time of all candidates in that bin. The weights are defined as the product of a background-subtraction weight and a kinematic correction weight which will be defined in the following paragraph. The gradient of the time-dependent asymmetry is determined from a linear least-squares fit to the measured asymmetries.

\begin{figure}[htbp]
  \begin{center}
    \includegraphics[width=0.49\linewidth]{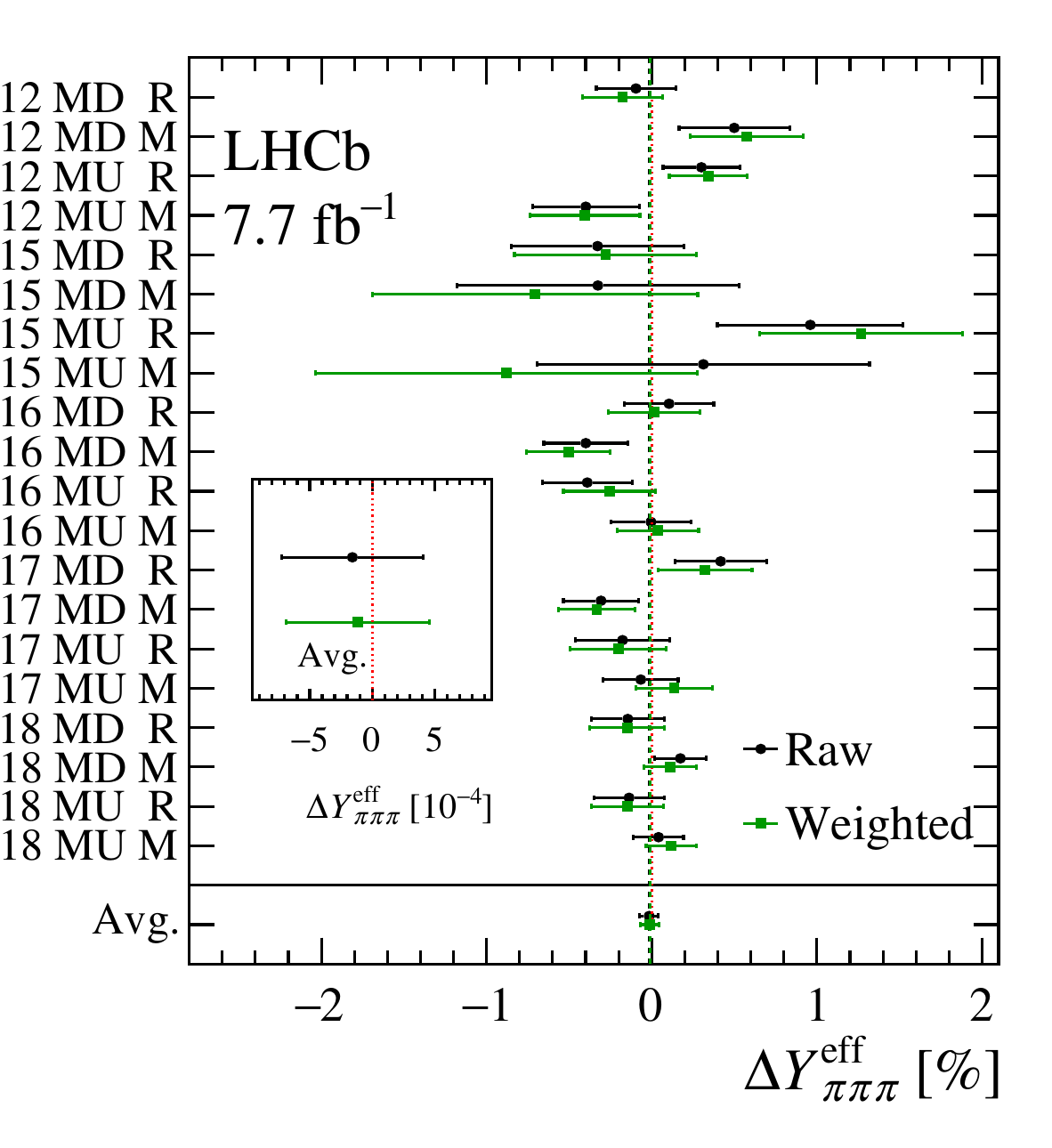}
    \includegraphics[width=0.49\linewidth]{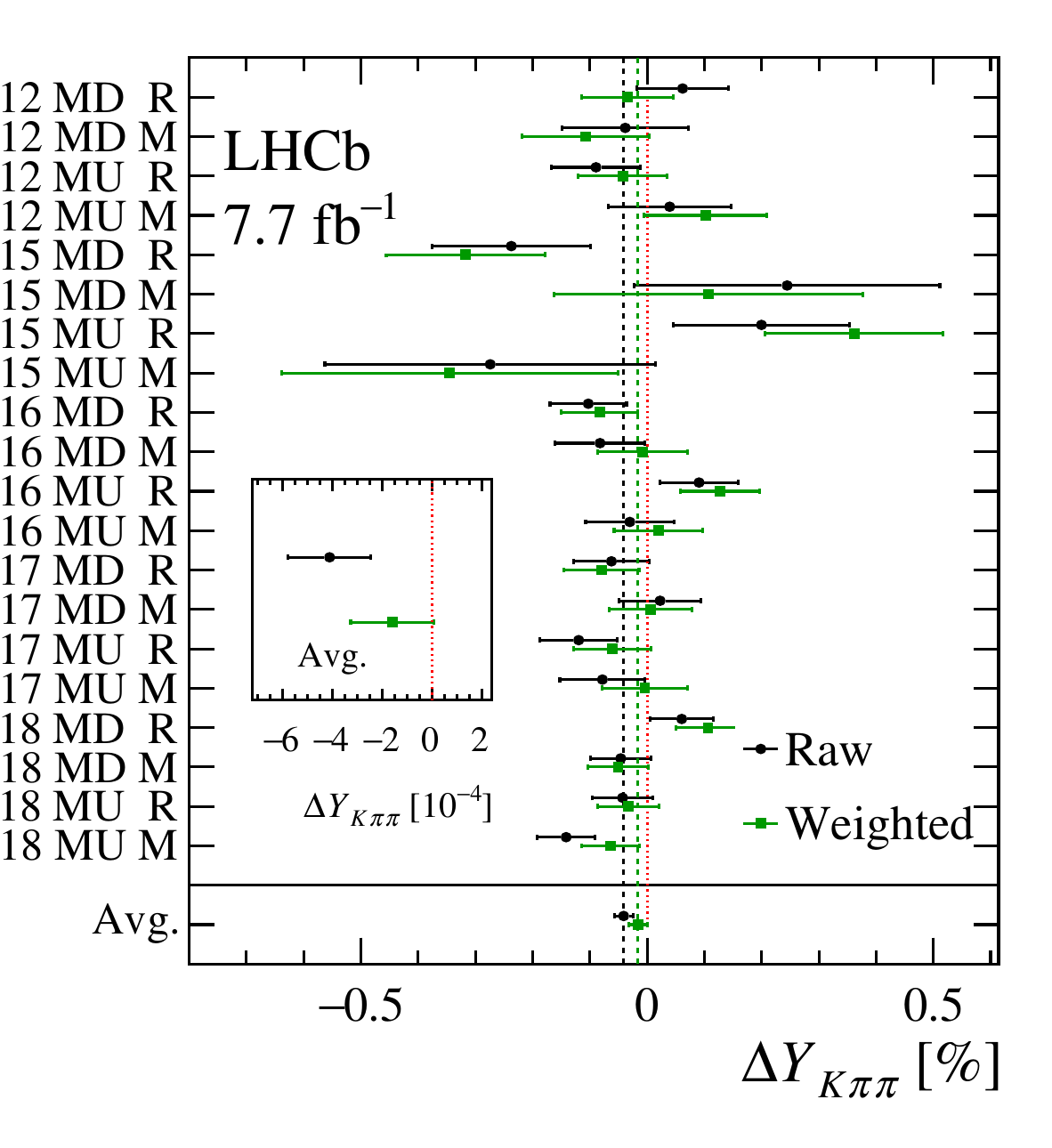}
    \vspace*{-0.5cm}
  \end{center}
  \caption{\small Measured gradient of the time-dependent asymmetry before and after kinematic weighting in each of the subsamples for (left) the signal channel and (right) the control channel. The abbreviation Avg. denotes the weighted average across all datasets obtained from a least squares fit, MU (MD) denotes the \MagUp (\MagDown) magnet polarity and M (R) denotes the merged (resolved) $\pi^0$ category. The black and green dotted lines represent the average measured gradients before and after the kinematic weighting, respectively, and the red dotted line indicates a value of zero. Only statistical uncertainties are shown.}
  \label{fig:deltay-combination}
\end{figure}

The raw asymmetries --- defined as the unweighted asymmetries between \Dz and \Dzb yields --- present in each data sample do not correspond to the \CP asymmetry in Equation~\ref{eq:acp_t}. Rather, they include the production asymmetry of \Dstarpm mesons in \proton\proton collisions, which potentially carries a kinematic dependence~\cite{Norrbin:2000zc}, and some unknown nuisance asymmetries induced by the detection, reconstruction and selection procedure. In particular, the \lhcb magnet deflects oppositely-charged tag pions in opposite directions, which induces some kinematic-dependent asymmetries, since the detector is not perfectly symmetrical. Periodically reversing the magnetic field polarity throughout the data-taking only approximately cancels the effect. Such kinematic asymmetries can introduce a time-dependent asymmetry in the presence of correlations between kinematic variables and the decay time of each \Dz candidate induced by the selection criteria. A per-event weighting procedure is applied to correct for the most severe kinematic-dependent asymmetries. Events are weighted, in several stages, to equalise the binned distributions of the curvature and projection angles of the tag-pion track~\cite{LHCb-PAPER-2020-045}, the pseudorapidity of the tag pion and \Dz candidates, and the transverse momentum of the \Dz candidate. Figure~\ref{fig:deltay-combination} shows the measured gradient of the time-dependent asymmetry in each data subsample before and after the kinematic weighting. No significant shift is observed in the signal channel. The measured value of $\Delta Y_{K\pi\pi}$ is compatible with zero at the level of $2.5\sigma$ before the kinematic weighting and $1.0\sigma$ after the kinematic weighting, where $\sigma$ is the statistical uncertainty. Finally, a set of pseudoexperiments is performed to evaluate and correct for the dilution of the measured $\Delta Y$ caused by the kinematic weighting, since a true time-dependent asymmetry would induce some kinematic-dependent asymmetries through the aforementioned correlations.

The systematic uncertainties associated with the measured time-dependent asymmetries are given in Table~\ref{tab:systematics} for both the signal and control modes. The dominant systematic uncertainties arise from possible kinematic-dependent detection asymmetries affecting the charged products of the \Dz meson decay. Potential biases from such sources are assessed by measuring the raw asymmetry as a function of phase-space position and decay time using the decays $\decay{D^+_{(s)}}{\pip\pip\pim}$, $\decay{D^+_{(s)}}{\phi(\decay{\, }{\Km\Kp})\pip}$, and $\decay{\Dp}{\Km\pip\pip}$ as~\cite{LHCb-PAPER-2021-009, LHCb-PAPER-2022-024}
\begin{equation}
    \begin{aligned}
        A_\mathrm{det}^{\pi\pi} = A_{\decay{D^+_{(s)}}{\pip\pip\pim}} &- A_{\decay{D^+_{(s)}}{\phi(\decay{\, }{\Km\Kp})\pip}}, \\
        A_\mathrm{det}^{K\pi} = A_{\decay{\Dp}{\Km\pip\pip}} &- A_{\decay{\Dp}{\phi(\decay{\, }{\Km\Kp})\pip}},
    \end{aligned}
\end{equation}
where $A_\mathrm{det}^{\pi\pi}$ ($A_\mathrm{det}^{K\pi}$) represents the detection asymmetry between a \pip\pim (\Km\pip) pair and its charge-conjugate at a given phase-space position and decay time. In each of these calibration decays, candidates are weighted so that the kinematic distributions agree between each pair of $D^+_{(s)}$ decays. This ensures a precise cancellation of kinematic-dependent detection and production asymmetries affecting the $D_{(s)}^+$ meson and the additional charged pion. Candidates are further weighted to align the kinematic distributions with those observed in the relevant \Dz decay mode, to accurately reproduce any kinematic-dependent \pip\pim or \Km\pip detection asymmetries~\cite{LHCb-PAPER-2021-009}. Potential direct \CP asymmetries in the calibration modes do not induce kinematic or decay-time dependent effects and are therefore irrelevant in this procedure. The measured asymmetry maps are then used as inputs for a set of pseudoexperiments to determine the potential biases on $\Delta Y_{\pi\pi\pi}^\text{eff}$ and $\Delta Y_{K\pi\pi}$.

\begin{table}[tb]
    \centering
    \caption{\small Systematic uncertainties affecting $\Delta Y^\text{eff}_{\pi\pi\pi}$ ($\Delta Y_{K\pi\pi}$) in the signal (control) channel.}
    \renewcommand{\arraystretch}{1.15}
    \begin{tabular}{ccc}
         \hline\hline
         Source & $\Delta Y_{\pi\pi\pi}^\text{eff}$ ($10^{-4}$) & $\Delta Y_{K\pi\pi}$ ($10^{-4}$) \\
         \hline
         Detection asymmetries & 1.6 & 3.4 \\
         $t/\tau_\Dz$ binning & 1.0 & 0.14 \\
         Secondary contamination & 0.84 & 0.84 \\
         \dm fit model & 0.75 & 0.08\\
         Kinematic weighting & 0.22 & 0.22\\
         \hline
         Total & 2.3 & 3.5\\
         \hline\hline
    \end{tabular}
    \label{tab:systematics}
\end{table}

A systematic uncertainty on the effect of the decay-time binning is conservatively assessed by performing the analysis with different choices of the decay-time binning scheme. The fraction of contamination from secondary decays and their asymmetries are measured in each decay-time bin by fitting the distribution of $\ln\chisqip$ for selected candidates using templates from simulation and the \dm sideband data as input. The measured fractions and asymmetries are then used in a pseudoexperiment study to determine the resulting systematic uncertainty. An average secondary fraction at the percent level is observed, depending on the data-taking period and \piz candidate reconstruction category. The uncertainty arising from the $\Delta m$ fit model is assessed in a set of pseudoexperiments using an alternative fit model to describe the signal and background \dm distributions. The number of bins used in the kinematic weighting is varied between a factor of one half and two, and the resulting standard deviation of measured asymmetry gradients is taken as a systematic uncertainty. A set of pseudoexperiments is carried out to determine the dilution of the measured asymmetry due to the finite decay-time resolution. The differences between the measured gradients with and without correcting for this effect are found to be negligible. Finally, the uncertainty on the measured $\Delta Y_{\pi\pi\pi}^\text{eff}$ related to the externally measured \Dz lifetime~\cite{PDG2022, Belle-II:2021cxx} is found to be negligible. No statistically significant effects are observed when stability checks are performed by applying a hardware-trigger requirement or adopting different strategies for the kinematic weighting and yield determination. Systematic uncertainties associated with the knowledge of the externally measured \CP-even fraction and the effect of a nonuniform phase-space acceptance on its central value are found to be negligible.

\begin{figure}[tbp]
  \begin{center}
    \includegraphics[width=0.49\linewidth]{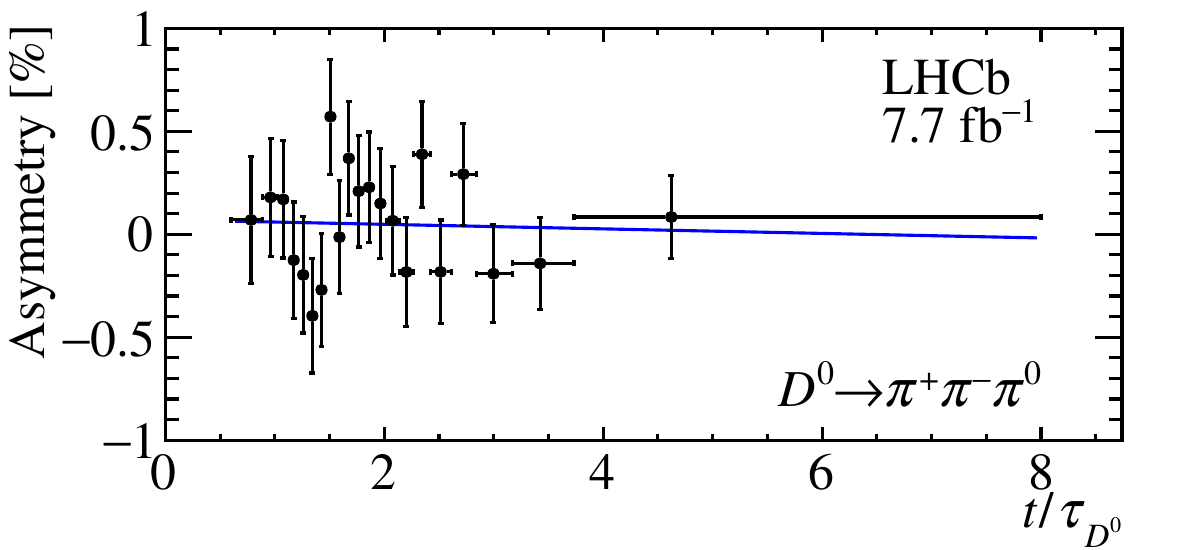}
    \includegraphics[width=0.49\linewidth]{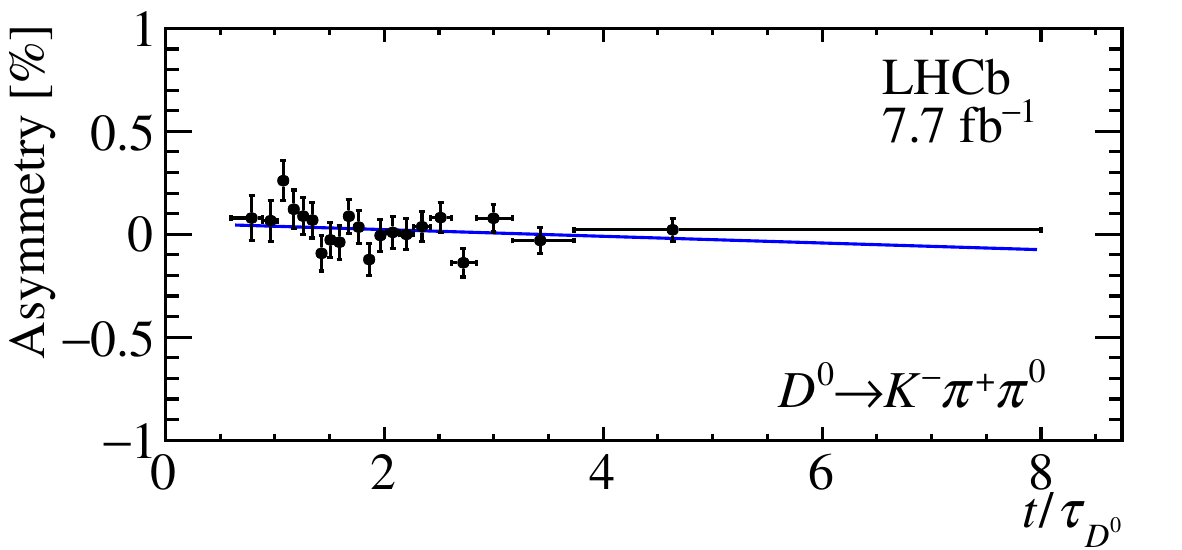}
    \vspace*{-0.5cm}
  \end{center}
  \caption{\small Measured asymmetry of (left) \decay{\Dz}{\pip\pim\piz} and (right) \decay{\Dz}{\Km\pip\piz} decays as a function of decay time, after kinematic weighting. The \chisq per degree of freedom of the linear fits shown alongside the data are $17/19$ and $22/19$, respectively.}
  \label{fig:slopes}
\end{figure}

By performing a least-squares fit to the measured asymmetry gradients in all subsets of the data, the final result \mbox{$\Delta Y_{\pi\pi\pi}^\text{eff} = (-1.2 \pm 6.0 \pm 2.3) \times 10^{-4}$} is obtained. The first uncertainty is statistical and the second is systematic, here and in the following. No evidence for time-dependent \CP violation in this channel is found. This corresponds, using the externally measured \CP-even fraction~\cite{Malde:2015mha}, to
\begin{equation*}
    \Delta Y = (-1.3 \pm 6.3 \pm 2.4) \times 10^{-4}.
\end{equation*}
This result is in excellent agreement with the current world average~\cite{HFLAV21}. In the control channel, the measurement \mbox{$\Delta Y_{K\pi\pi} = (-1.7 \pm 1.8 \pm 3.5) \times 10^{-4}$} is obtained, which validates the kinematic-weighting procedure designed to account for nuisance asymmetries. Figure~\ref{fig:slopes} shows the measured asymmetry for all selected \decay{\Dz}{\pip\pim\piz} and \decay{\Dz}{\Km\pip\piz} candidates as a function of decay time, along with a linear fit.

In summary, a measurement of time-dependent \CP violation in the decay \decay{\Dz}{\pip\pim\piz} has been presented. No evidence for \CP violation is found. The precision of this measurement is limited by the statistical uncertainty and the result is in excellent agreement with the current world average of the parameter $\Delta Y$. This represents the first measurement of time-dependent \CP violation in the singly Cabibbo-suppressed decay \decay{\Dz}{\pip\pim\piz}.

\section*{Acknowledgements}
%
%
\noindent We express our gratitude to our colleagues in the CERN
accelerator departments for the excellent performance of the LHC. We
thank the technical and administrative staff at the LHCb
institutes.
We acknowledge support from CERN and from the national agencies:
CAPES, CNPq, FAPERJ and FINEP (Brazil); 
MOST and NSFC (China); 
CNRS/IN2P3 (France); 
BMBF, DFG and MPG (Germany); 
INFN (Italy); 
NWO (Netherlands); 
MNiSW and NCN (Poland); 
MCID/IFA (Romania); 
MICINN (Spain); 
SNSF and SER (Switzerland); 
NASU (Ukraine); 
STFC (United Kingdom); 
DOE NP and NSF (USA).
We acknowledge the computing resources that are provided by CERN, IN2P3
(France), KIT and DESY (Germany), INFN (Italy), SURF (Netherlands),
PIC (Spain), GridPP (United Kingdom), 
CSCS (Switzerland), IFIN-HH (Romania), CBPF (Brazil),
and Polish WLCG (Poland).
We are indebted to the communities behind the multiple open-source
software packages on which we depend.
Individual groups or members have received support from
ARC and ARDC (Australia);
Key Research Program of Frontier Sciences of CAS, CAS PIFI, CAS CCEPP, 
Fundamental Research Funds for the Central Universities, 
and Sci. \& Tech. Program of Guangzhou (China);
Minciencias (Colombia);
EPLANET, Marie Sk\l{}odowska-Curie Actions, ERC and NextGenerationEU (European Union);
A*MIDEX, ANR, IPhU and Labex P2IO, and R\'{e}gion Auvergne-Rh\^{o}ne-Alpes (France);
AvH Foundation (Germany);
ICSC (Italy); 
GVA, XuntaGal, GENCAT, Inditex, InTalent and Prog.~Atracci\'on Talento, CM (Spain);
SRC (Sweden);
the Leverhulme Trust, the Royal Society
 and UKRI (United Kingdom).




\addcontentsline{toc}{section}{References}
\bibliographystyle{LHCb}
\bibliography{main,standard,LHCb-PAPER,LHCb-CONF,LHCb-DP,LHCb-TDR}

\ifx\mcitethebibliography\mciteundefinedmacro
\PackageError{LHCb.bst}{mciteplus.sty has not been loaded}
{This bibstyle requires the use of the mciteplus package.}\fi
\providecommand{\href}[2]{#2}
\begin{mcitethebibliography}{10}
\mciteSetBstSublistMode{n}
\mciteSetBstMaxWidthForm{subitem}{\alph{mcitesubitemcount})}
\mciteSetBstSublistLabelBeginEnd{\mcitemaxwidthsubitemform\space}
{\relax}{\relax}

\bibitem{Sakharov:1967dj}
A.~D. Sakharov, \ifthenelse{\boolean{articletitles}}{\emph{{Violation of CP invariance, C asymmetry, and baryon asymmetry of the universe}}, }{}\href{https://doi.org/10.1070/PU1991v034n05ABEH002497}{Pisma Zh.\ Eksp.\ Teor.\ Fiz.\  \textbf{5} (1967) 32}\relax
\mciteBstWouldAddEndPuncttrue
\mciteSetBstMidEndSepPunct{\mcitedefaultmidpunct}
{\mcitedefaultendpunct}{\mcitedefaultseppunct}\relax
\EndOfBibitem
\bibitem{Dine:2003ax}
M.~Dine and A.~Kusenko, \ifthenelse{\boolean{articletitles}}{\emph{{Origin of the matter - antimatter asymmetry}}, }{}\href{https://doi.org/10.1103/RevModPhys.76.1}{Rev.\ Mod.\ Phys.\  \textbf{76} (2003) 1}, \href{http://arxiv.org/abs/hep-ph/0303065}{{\normalfont\ttfamily arXiv:hep-ph/0303065}}\relax
\mciteBstWouldAddEndPuncttrue
\mciteSetBstMidEndSepPunct{\mcitedefaultmidpunct}
{\mcitedefaultendpunct}{\mcitedefaultseppunct}\relax
\EndOfBibitem
\bibitem{Canetti:2012zc}
L.~Canetti, M.~Drewes, and M.~Shaposhnikov, \ifthenelse{\boolean{articletitles}}{\emph{{Matter and antimatter in the universe}}, }{}\href{https://doi.org/10.1088/1367-2630/14/9/095012}{New J.\ Phys.\  \textbf{14} (2012) 095012}, \href{http://arxiv.org/abs/1204.4186}{{\normalfont\ttfamily arXiv:1204.4186}}\relax
\mciteBstWouldAddEndPuncttrue
\mciteSetBstMidEndSepPunct{\mcitedefaultmidpunct}
{\mcitedefaultendpunct}{\mcitedefaultseppunct}\relax
\EndOfBibitem
\bibitem{Christenson:1964fg}
J.~H. Christenson, J.~W. Cronin, V.~L. Fitch, and R.~Turlay, \ifthenelse{\boolean{articletitles}}{\emph{{Evidence for the $2\pi$ decay of the $K_2^0$ meson}}, }{}\href{https://doi.org/10.1103/PhysRevLett.13.138}{Phys.\ Rev.\ Lett.\  \textbf{13} (1964) 138}\relax
\mciteBstWouldAddEndPuncttrue
\mciteSetBstMidEndSepPunct{\mcitedefaultmidpunct}
{\mcitedefaultendpunct}{\mcitedefaultseppunct}\relax
\EndOfBibitem
\bibitem{KTeV:1999kad}
KTeV collaboration, A.~Alavi-Harati {\em et~al.}, \ifthenelse{\boolean{articletitles}}{\emph{{Observation of direct CP violation in $K_\mathrm{S,L} \to \pi \pi$ decays}}, }{}\href{https://doi.org/10.1103/PhysRevLett.83.22}{Phys.\ Rev.\ Lett.\  \textbf{83} (1999) 22}, \href{http://arxiv.org/abs/hep-ex/9905060}{{\normalfont\ttfamily arXiv:hep-ex/9905060}}\relax
\mciteBstWouldAddEndPuncttrue
\mciteSetBstMidEndSepPunct{\mcitedefaultmidpunct}
{\mcitedefaultendpunct}{\mcitedefaultseppunct}\relax
\EndOfBibitem
\bibitem{BaBar:2001pki}
BaBar collaboration, B.~Aubert {\em et~al.}, \ifthenelse{\boolean{articletitles}}{\emph{{Observation of CP violation in the $B^0$ meson system}}, }{}\href{https://doi.org/10.1103/PhysRevLett.87.091801}{Phys.\ Rev.\ Lett.\  \textbf{87} (2001) 091801}, \href{http://arxiv.org/abs/hep-ex/0107013}{{\normalfont\ttfamily arXiv:hep-ex/0107013}}\relax
\mciteBstWouldAddEndPuncttrue
\mciteSetBstMidEndSepPunct{\mcitedefaultmidpunct}
{\mcitedefaultendpunct}{\mcitedefaultseppunct}\relax
\EndOfBibitem
\bibitem{Belle:2001zzw}
Belle collaboration, K.~Abe {\em et~al.}, \ifthenelse{\boolean{articletitles}}{\emph{{Observation of large CP violation in the neutral $B$ meson system}}, }{}\href{https://doi.org/10.1103/PhysRevLett.87.091802}{Phys.\ Rev.\ Lett.\  \textbf{87} (2001) 091802}, \href{http://arxiv.org/abs/hep-ex/0107061}{{\normalfont\ttfamily arXiv:hep-ex/0107061}}\relax
\mciteBstWouldAddEndPuncttrue
\mciteSetBstMidEndSepPunct{\mcitedefaultmidpunct}
{\mcitedefaultendpunct}{\mcitedefaultseppunct}\relax
\EndOfBibitem
\bibitem{LHCB-PAPER-2013-018}
LHCb collaboration, R.~Aaij {\em et~al.}, \ifthenelse{\boolean{articletitles}}{\emph{{First observation of \CP violation in the decays of \Bs mesons}}, }{}\href{https://doi.org/10.1103/PhysRevLett.110.221601}{Phys.\ Rev.\ Lett.\  \textbf{110} (2013) 221601}, \href{http://arxiv.org/abs/1304.6173}{{\normalfont\ttfamily arXiv:1304.6173}}\relax
\mciteBstWouldAddEndPuncttrue
\mciteSetBstMidEndSepPunct{\mcitedefaultmidpunct}
{\mcitedefaultendpunct}{\mcitedefaultseppunct}\relax
\EndOfBibitem
\bibitem{LHCb-PAPER-2020-029}
LHCb collaboration, R.~Aaij {\em et~al.}, \ifthenelse{\boolean{articletitles}}{\emph{{Observation of \CP violation in two-body \mbox{$\BdorBs$-meson} decays to charged pions and kaons}}, }{}\href{https://doi.org/10.1007/JHEP03(2021)075}{JHEP \textbf{03} (2021) 075}, \href{http://arxiv.org/abs/2012.05319}{{\normalfont\ttfamily arXiv:2012.05319}}\relax
\mciteBstWouldAddEndPuncttrue
\mciteSetBstMidEndSepPunct{\mcitedefaultmidpunct}
{\mcitedefaultendpunct}{\mcitedefaultseppunct}\relax
\EndOfBibitem
\bibitem{LHCb-PAPER-2019-006}
LHCb collaboration, R.~Aaij {\em et~al.}, \ifthenelse{\boolean{articletitles}}{\emph{{Observation of \CP violation in charm decays}}, }{}\href{https://doi.org/10.1103/PhysRevLett.122.211803}{Phys.\ Rev.\ Lett.\  \textbf{122} (2019) 211803}, \href{http://arxiv.org/abs/1903.08726}{{\normalfont\ttfamily arXiv:1903.08726}}\relax
\mciteBstWouldAddEndPuncttrue
\mciteSetBstMidEndSepPunct{\mcitedefaultmidpunct}
{\mcitedefaultendpunct}{\mcitedefaultseppunct}\relax
\EndOfBibitem
\bibitem{LHCb-PAPER-2022-024}
LHCb collaboration, R.~Aaij {\em et~al.}, \ifthenelse{\boolean{articletitles}}{\emph{{Measurement of the time-integrated \CP asymmetry in $\Dz \to \Km \Kp$ decays}}, }{}\href{https://doi.org/10.1103/PhysRevLett.131.091802}{Phys.\ Rev.\ Lett.\  \textbf{131} (2023) 091802}, \href{http://arxiv.org/abs/2209.03179}{{\normalfont\ttfamily arXiv:2209.03179}}\relax
\mciteBstWouldAddEndPuncttrue
\mciteSetBstMidEndSepPunct{\mcitedefaultmidpunct}
{\mcitedefaultendpunct}{\mcitedefaultseppunct}\relax
\EndOfBibitem
\bibitem{HFLAV21}
Y.~Amhis {\em et~al.}, \ifthenelse{\boolean{articletitles}}{\emph{{Averages of $b$-hadron, $c$-hadron, and $\tau$-lepton properties as of 2021}}, }{}\href{https://doi.org/10.1103/PhysRevD.107.052008}{Phys.\ Rev.\  \textbf{D107} (2023) 052008}, \href{http://arxiv.org/abs/2206.07501}{{\normalfont\ttfamily arXiv:2206.07501}}, {updated results and plots available at \href{https://hflav.web.cern.ch}{{\texttt{https://hflav.web.cern.ch}}}}\relax
\mciteBstWouldAddEndPuncttrue
\mciteSetBstMidEndSepPunct{\mcitedefaultmidpunct}
{\mcitedefaultendpunct}{\mcitedefaultseppunct}\relax
\EndOfBibitem
\bibitem{LHCb-PAPER-2016-032}
LHCb collaboration, R.~Aaij {\em et~al.}, \ifthenelse{\boolean{articletitles}}{\emph{{Measurement of the CKM angle $\gamma$ from a combination of LHCb results}}, }{}\href{https://doi.org/10.1007/JHEP12(2016)087}{JHEP \textbf{12} (2016) 087}, \href{http://arxiv.org/abs/1611.03076}{{\normalfont\ttfamily arXiv:1611.03076}}\relax
\mciteBstWouldAddEndPuncttrue
\mciteSetBstMidEndSepPunct{\mcitedefaultmidpunct}
{\mcitedefaultendpunct}{\mcitedefaultseppunct}\relax
\EndOfBibitem
\bibitem{GammaCombo}
M.~Kenzie {\em et~al.}, \ifthenelse{\boolean{articletitles}}{\emph{{GammaCombo: A statistical analysis framework for combining measurements, fitting datasets and producing confidence intervals}}, }{}
\newblock doi:~\href{https://doi.org/10.5281/zenodo.3371421}{10.5281/zenodo.3371421}\relax
\mciteBstWouldAddEndPuncttrue
\mciteSetBstMidEndSepPunct{\mcitedefaultmidpunct}
{\mcitedefaultendpunct}{\mcitedefaultseppunct}\relax
\EndOfBibitem
\bibitem{PDG2022}
Particle Data Group, R.~L. Workman {\em et~al.}, \ifthenelse{\boolean{articletitles}}{\emph{{\href{http://pdg.lbl.gov/}{Review of particle physics}}}, }{}\href{https://doi.org/10.1093/ptep/ptac097}{Prog.\ Theor.\ Exp.\ Phys.\  \textbf{2022} (2022) 083C01}\relax
\mciteBstWouldAddEndPuncttrue
\mciteSetBstMidEndSepPunct{\mcitedefaultmidpunct}
{\mcitedefaultendpunct}{\mcitedefaultseppunct}\relax
\EndOfBibitem
\bibitem{Belle-II:2021cxx}
Belle-II collaboration, F.~Abudin\'en {\em et~al.}, \ifthenelse{\boolean{articletitles}}{\emph{{Precise measurement of the $D^0$ and $D^+$ lifetimes at Belle II}}, }{}\href{https://doi.org/10.1103/PhysRevLett.127.211801}{Phys.\ Rev.\ Lett.\  \textbf{127} (2021) 211801}, \href{http://arxiv.org/abs/2108.03216}{{\normalfont\ttfamily arXiv:2108.03216}}\relax
\mciteBstWouldAddEndPuncttrue
\mciteSetBstMidEndSepPunct{\mcitedefaultmidpunct}
{\mcitedefaultendpunct}{\mcitedefaultseppunct}\relax
\EndOfBibitem
\bibitem{Kagan:2020vri}
A.~L. Kagan and L.~Silvestrini, \ifthenelse{\boolean{articletitles}}{\emph{{Dispersive and absorptive $CP$ violation in $\Dz - \Dzb$ mixing}}, }{}\href{https://doi.org/10.1103/PhysRevD.103.053008}{Phys.\ Rev.\  \textbf{D103} (2021) 053008}, \href{http://arxiv.org/abs/2001.07207}{{\normalfont\ttfamily arXiv:2001.07207}}\relax
\mciteBstWouldAddEndPuncttrue
\mciteSetBstMidEndSepPunct{\mcitedefaultmidpunct}
{\mcitedefaultendpunct}{\mcitedefaultseppunct}\relax
\EndOfBibitem
\bibitem{BaBar:2012bho}
BaBar collaboration, J.~P. Lees {\em et~al.}, \ifthenelse{\boolean{articletitles}}{\emph{{Measurement of $\Dz-\Dzb$ mixing and CP violation in two-body $D^0$ Decays}}, }{}\href{https://doi.org/10.1103/PhysRevD.87.012004}{Phys.\ Rev.\  \textbf{D87} (2013) 012004}, \href{http://arxiv.org/abs/1209.3896}{{\normalfont\ttfamily arXiv:1209.3896}}\relax
\mciteBstWouldAddEndPuncttrue
\mciteSetBstMidEndSepPunct{\mcitedefaultmidpunct}
{\mcitedefaultendpunct}{\mcitedefaultseppunct}\relax
\EndOfBibitem
\bibitem{CDF:2014wyb}
CDF collaboration, T.~A. Aaltonen {\em et~al.}, \ifthenelse{\boolean{articletitles}}{\emph{{Measurement of indirect CP-violating asymmetries in $D^0\to K^+K^-$ and $D^0\to \pi^+\pi^-$ decays at CDF}}, }{}\href{https://doi.org/10.1103/PhysRevD.90.111103}{Phys.\ Rev.\  \textbf{D90} (2014) 111103}, \href{http://arxiv.org/abs/1410.5435}{{\normalfont\ttfamily arXiv:1410.5435}}\relax
\mciteBstWouldAddEndPuncttrue
\mciteSetBstMidEndSepPunct{\mcitedefaultmidpunct}
{\mcitedefaultendpunct}{\mcitedefaultseppunct}\relax
\EndOfBibitem
\bibitem{Belle:2015etc}
Belle collaboration, M.~Stari\v{c} {\em et~al.}, \ifthenelse{\boolean{articletitles}}{\emph{{Measurement of $\Dz-\Dzb$ mixing and search for CP violation in $D^0 \to K^+K^-, \pi^+\pi^-$ decays with the full Belle data set}}, }{}\href{https://doi.org/10.1016/j.physletb.2015.12.025}{Phys.\ Lett.\  \textbf{B753} (2016) 412}, \href{http://arxiv.org/abs/1509.08266}{{\normalfont\ttfamily arXiv:1509.08266}}\relax
\mciteBstWouldAddEndPuncttrue
\mciteSetBstMidEndSepPunct{\mcitedefaultmidpunct}
{\mcitedefaultendpunct}{\mcitedefaultseppunct}\relax
\EndOfBibitem
\bibitem{LHCb-PAPER-2014-069}
LHCb collaboration, R.~Aaij {\em et~al.}, \ifthenelse{\boolean{articletitles}}{\emph{{Measurement of indirect \CP asymmetries in \mbox{\decay{\Dz}{\Km\Kp}} and \mbox{\decay{\Dz}{\pim\pip}} decays using semileptonic \B decays}}, }{}\href{https://doi.org/10.1007/JHEP04(2015)043}{JHEP \textbf{04} (2015) 043}, \href{http://arxiv.org/abs/1501.06777}{{\normalfont\ttfamily arXiv:1501.06777}}\relax
\mciteBstWouldAddEndPuncttrue
\mciteSetBstMidEndSepPunct{\mcitedefaultmidpunct}
{\mcitedefaultendpunct}{\mcitedefaultseppunct}\relax
\EndOfBibitem
\bibitem{LHCb-PAPER-2016-063}
LHCb collaboration, R.~Aaij {\em et~al.}, \ifthenelse{\boolean{articletitles}}{\emph{{Measurement of the \CP violation parameter $A_\Gamma$ in \mbox{\decay{\Dz}{\Kp\Km}} and \mbox{\decay{\Dz}{\pip\pim}} decays}}, }{}\href{https://doi.org/10.1103/PhysRevLett.118.261803}{Phys.\ Rev.\ Lett.\  \textbf{118} (2017) 261803}, \href{http://arxiv.org/abs/1702.06490}{{\normalfont\ttfamily arXiv:1702.06490}}\relax
\mciteBstWouldAddEndPuncttrue
\mciteSetBstMidEndSepPunct{\mcitedefaultmidpunct}
{\mcitedefaultendpunct}{\mcitedefaultseppunct}\relax
\EndOfBibitem
\bibitem{LHCb-PAPER-2019-032}
LHCb collaboration, R.~Aaij {\em et~al.}, \ifthenelse{\boolean{articletitles}}{\emph{{Updated measurement of decay-time-dependent \CP asymmetries in \mbox{\decay{\Dz}{\Kp\Km}} and \mbox{\decay{\Dz}{\pip\pim}} decays}}, }{}\href{https://doi.org/10.1103/PhysRevD.101.012005}{Phys.\ Rev.\  \textbf{D101} (2020) 012005}, \href{http://arxiv.org/abs/1911.01114}{{\normalfont\ttfamily arXiv:1911.01114}}\relax
\mciteBstWouldAddEndPuncttrue
\mciteSetBstMidEndSepPunct{\mcitedefaultmidpunct}
{\mcitedefaultendpunct}{\mcitedefaultseppunct}\relax
\EndOfBibitem
\bibitem{LHCb-PAPER-2020-045}
LHCb collaboration, R.~Aaij {\em et~al.}, \ifthenelse{\boolean{articletitles}}{\emph{{Search for time-dependent \CP violation in \mbox{$\Dz \to \Kp \Km$} and \mbox{$\Dz \to \pip \pim$} decays}}, }{}\href{https://doi.org/10.1103/PhysRevD.104.072010}{Phys.\ Rev.\  \textbf{D104} (2021) 072010}, \href{http://arxiv.org/abs/2105.09889}{{\normalfont\ttfamily arXiv:2105.09889}}\relax
\mciteBstWouldAddEndPuncttrue
\mciteSetBstMidEndSepPunct{\mcitedefaultmidpunct}
{\mcitedefaultendpunct}{\mcitedefaultseppunct}\relax
\EndOfBibitem
\bibitem{LHCb-PAPER-2021-009}
LHCb collaboration, R.~Aaij {\em et~al.}, \ifthenelse{\boolean{articletitles}}{\emph{{Observation of the mass difference between neutral charm-meson eigenstates}}, }{}\href{https://doi.org/10.1103/PhysRevLett.127.111801}{Phys.\ Rev.\ Lett.\  \textbf{127} (2021) 111801}, Erratum \href{https://doi.org/10.1103/PhysRevLett.131.079901}{ibid.\   \textbf{131} (2023) 079901}, \href{http://arxiv.org/abs/2106.03744}{{\normalfont\ttfamily arXiv:2106.03744}}\relax
\mciteBstWouldAddEndPuncttrue
\mciteSetBstMidEndSepPunct{\mcitedefaultmidpunct}
{\mcitedefaultendpunct}{\mcitedefaultseppunct}\relax
\EndOfBibitem
\bibitem{LHCb-PAPER-2022-020}
LHCb collaboration, R.~Aaij {\em et~al.}, \ifthenelse{\boolean{articletitles}}{\emph{{Model-independent measurement of charm mixing parameters in $\Bbar \to \Dz (\to \Kstarz \pip \pim) \mun \neumb X$ decays}}, }{}\href{https://doi.org/10.1103/PhysRevD.108.052005}{Phys.\ Rev.\  \textbf{D108} (2023) 052005}, \href{http://arxiv.org/abs/2208.06512}{{\normalfont\ttfamily arXiv:2208.06512}}\relax
\mciteBstWouldAddEndPuncttrue
\mciteSetBstMidEndSepPunct{\mcitedefaultmidpunct}
{\mcitedefaultendpunct}{\mcitedefaultseppunct}\relax
\EndOfBibitem
\bibitem{Malde:2015xra}
S.~Malde, C.~Thomas, and G.~Wilkinson, \ifthenelse{\boolean{articletitles}}{\emph{{Measuring $CP$ violation and mixing in charm with inclusive self-conjugate multibody decay modes}}, }{}\href{https://doi.org/10.1103/PhysRevD.91.094032}{Phys.\ Rev.\  \textbf{D91} (2015) 094032}, \href{http://arxiv.org/abs/1502.04560}{{\normalfont\ttfamily arXiv:1502.04560}}\relax
\mciteBstWouldAddEndPuncttrue
\mciteSetBstMidEndSepPunct{\mcitedefaultmidpunct}
{\mcitedefaultendpunct}{\mcitedefaultseppunct}\relax
\EndOfBibitem
\bibitem{Malde:2015mha}
S.~Malde {\em et~al.}, \ifthenelse{\boolean{articletitles}}{\emph{{First determination of the $CP$ content of $D \to \pi^+\pi^-\pi^+\pi^-$ and updated determination of the $CP$ contents of $D \to \pi^+\pi^-\pi^0$ and $D \to K^+K^-\pi^0$}}, }{}\href{https://doi.org/10.1016/j.physletb.2015.05.043}{Phys.\ Lett.\  \textbf{B747} (2015) 9}, \href{http://arxiv.org/abs/1504.05878}{{\normalfont\ttfamily arXiv:1504.05878}}\relax
\mciteBstWouldAddEndPuncttrue
\mciteSetBstMidEndSepPunct{\mcitedefaultmidpunct}
{\mcitedefaultendpunct}{\mcitedefaultseppunct}\relax
\EndOfBibitem
\bibitem{LHCb-DP-2008-001}
LHCb collaboration, A.~A. Alves~Jr.\ {\em et~al.}, \ifthenelse{\boolean{articletitles}}{\emph{{The \lhcb detector at the LHC}}, }{}\href{https://doi.org/10.1088/1748-0221/3/08/S08005}{JINST \textbf{3} (2008) S08005}\relax
\mciteBstWouldAddEndPuncttrue
\mciteSetBstMidEndSepPunct{\mcitedefaultmidpunct}
{\mcitedefaultendpunct}{\mcitedefaultseppunct}\relax
\EndOfBibitem
\bibitem{LHCb-DP-2014-002}
LHCb collaboration, R.~Aaij {\em et~al.}, \ifthenelse{\boolean{articletitles}}{\emph{{LHCb detector performance}}, }{}\href{https://doi.org/10.1142/S0217751X15300227}{Int.\ J.\ Mod.\ Phys.\  \textbf{A30} (2015) 1530022}, \href{http://arxiv.org/abs/1412.6352}{{\normalfont\ttfamily arXiv:1412.6352}}\relax
\mciteBstWouldAddEndPuncttrue
\mciteSetBstMidEndSepPunct{\mcitedefaultmidpunct}
{\mcitedefaultendpunct}{\mcitedefaultseppunct}\relax
\EndOfBibitem
\bibitem{LHCb-PAPER-2014-054}
LHCb collaboration, R.~Aaij {\em et~al.}, \ifthenelse{\boolean{articletitles}}{\emph{{Search for \CP violation in \mbox{\decay{\Dz}{\pim\pip\piz}} decays with the energy test}}, }{}\href{https://doi.org/10.1016/j.physletb.2014.11.043}{Phys.\ Lett.\  \textbf{B740} (2015) 158}, \href{http://arxiv.org/abs/1410.4170}{{\normalfont\ttfamily arXiv:1410.4170}}\relax
\mciteBstWouldAddEndPuncttrue
\mciteSetBstMidEndSepPunct{\mcitedefaultmidpunct}
{\mcitedefaultendpunct}{\mcitedefaultseppunct}\relax
\EndOfBibitem
\bibitem{BBDT}
V.~V. Gligorov and M.~Williams, \ifthenelse{\boolean{articletitles}}{\emph{{Efficient, reliable and fast high-level triggering using a bonsai boosted decision tree}}, }{}\href{https://doi.org/10.1088/1748-0221/8/02/P02013}{JINST \textbf{8} (2013) P02013}, \href{http://arxiv.org/abs/1210.6861}{{\normalfont\ttfamily arXiv:1210.6861}}\relax
\mciteBstWouldAddEndPuncttrue
\mciteSetBstMidEndSepPunct{\mcitedefaultmidpunct}
{\mcitedefaultendpunct}{\mcitedefaultseppunct}\relax
\EndOfBibitem
\bibitem{Hulsbergen:2005pu}
W.~D. Hulsbergen, \ifthenelse{\boolean{articletitles}}{\emph{{Decay chain fitting with a Kalman filter}}, }{}\href{https://doi.org/10.1016/j.nima.2005.06.078}{Nucl.\ Instrum.\ Meth.\  \textbf{A552} (2005) 566}, \href{http://arxiv.org/abs/physics/0503191}{{\normalfont\ttfamily arXiv:physics/0503191}}\relax
\mciteBstWouldAddEndPuncttrue
\mciteSetBstMidEndSepPunct{\mcitedefaultmidpunct}
{\mcitedefaultendpunct}{\mcitedefaultseppunct}\relax
\EndOfBibitem
\bibitem{ROOT}
R.~Brun and F.~Rademakers, \ifthenelse{\boolean{articletitles}}{\emph{{ROOT: An object oriented data analysis framework}}, }{}\href{https://doi.org/10.1016/S0168-9002(97)00048-X}{Nucl.\ Instrum.\ Meth.\  \textbf{A389} (1997) 81}\relax
\mciteBstWouldAddEndPuncttrue
\mciteSetBstMidEndSepPunct{\mcitedefaultmidpunct}
{\mcitedefaultendpunct}{\mcitedefaultseppunct}\relax
\EndOfBibitem
\bibitem{RooFit}
W.~Verkerke and D.~P. Kirkby, \ifthenelse{\boolean{articletitles}}{\emph{{The RooFit toolkit for data modeling}}, }{}eConf \textbf{C0303241} (2003) MOLT007, \href{http://arxiv.org/abs/physics/0306116}{{\normalfont\ttfamily arXiv:physics/0306116}}\relax
\mciteBstWouldAddEndPuncttrue
\mciteSetBstMidEndSepPunct{\mcitedefaultmidpunct}
{\mcitedefaultendpunct}{\mcitedefaultseppunct}\relax
\EndOfBibitem
\bibitem{Norrbin:2000zc}
E.~Norrbin and T.~Sj{\"o}strand, \ifthenelse{\boolean{articletitles}}{\emph{{Production and hadronization of heavy quarks}}, }{}\href{https://doi.org/10.1007/s100520000460}{Eur.\ Phys.\ J.\ C \textbf{17} (2000) 137}, \href{http://arxiv.org/abs/hep-ph/0005110}{{\normalfont\ttfamily arXiv:hep-ph/0005110}}\relax
\mciteBstWouldAddEndPuncttrue
\mciteSetBstMidEndSepPunct{\mcitedefaultmidpunct}
{\mcitedefaultendpunct}{\mcitedefaultseppunct}\relax
\EndOfBibitem
\end{mcitethebibliography}

\newpage
\centerline
{\large\bf LHCb collaboration}
\begin
{flushleft}
\small
R.~Aaij$^{36}$\lhcborcid{0000-0003-0533-1952},
A.S.W.~Abdelmotteleb$^{55}$\lhcborcid{0000-0001-7905-0542},
C.~Abellan~Beteta$^{49}$,
F.~Abudin{\'e}n$^{55}$\lhcborcid{0000-0002-6737-3528},
T.~Ackernley$^{59}$\lhcborcid{0000-0002-5951-3498},
A. A. ~Adefisoye$^{67}$\lhcborcid{0000-0003-2448-1550},
B.~Adeva$^{45}$\lhcborcid{0000-0001-9756-3712},
M.~Adinolfi$^{53}$\lhcborcid{0000-0002-1326-1264},
P.~Adlarson$^{79}$\lhcborcid{0000-0001-6280-3851},
C.~Agapopoulou$^{47}$\lhcborcid{0000-0002-2368-0147}
% ...

{\footnotesize \it

% ...
$^{36}$Nikhef National Institute for Subatomic Physics, Amsterdam, Netherlands\\
% ...
$^{45}$Instituto Galego de F{\'\i}sica de Altas Enerx{\'\i}as (IGFAE), Universidade de Santiago de Compostela, Santiago de Compostela, Spain\\
% ...
$^{47}$European Organization for Nuclear Research (CERN), Geneva, Switzerland\\
% ...
$^{49}$Physik-Institut, Universit{\"a}t Z{\"u}rich, Z{\"u}rich, Switzerland\\
% ...
$^{53}$H.H. Wills Physics Laboratory, University of Bristol, Bristol, United Kingdom\\
% ...
$^{55}$Department of Physics, University of Warwick, Coventry, United Kingdom\\
% ...
$^{58}$School of Physics and Astronomy, University of Glasgow, Glasgow, United Kingdom\\
$^{59}$Oliver Lodge Laboratory, University of Liverpool, Liverpool, United Kingdom\\
% ...
$^{67}$Syracuse University, Syracuse, NY, United States\\
% ...
$^{79}$Department of Physics and Astronomy, Uppsala University, Uppsala, Sweden, associated to $^{58}$\\
% ...
}
\end{flushleft}

\end{document}